\documentclass[nonacm,acmsmall]{acmart}

\title{Trace Mutation in Human-LLM Dialogue}
\subtitle{The Transcript as Forensic and Mitigation Surface}
\usepackage{comment}
\usepackage{makecell}

\renewcommand\footnotetextcopyrightpermission[1]{}
\usepackage[table]{xcolor}
\hypersetup{%
  pdftitle={Trace Mutation in Human-LLM Dialogue: The Transcript as Forensic and Mitigation Surface},
  pdfauthor={William Bensen},
  pdflang={en},
  pdfkeywords={human--AI interaction, large language models, conversation analysis, grounding and repair}
}
\author{William Bensen}
\email{william.bensen.do@gmail.com}
\orcid{0009-0007-2283-4937}
\affiliation{%
  \institution{Independent Researcher}
  \city{Spencer}
  \state{Iowa}
  \country{USA}
}

\usepackage{float}

\begin{abstract}
Large language models (LLMs) are increasingly deployed as partners in knowledge work, where the shared conversational record functions as the decision record that safeguards work continuity. We characterize a class of context failures we term \emph{trace mutations}, in which distortions enter the shared record while presenting as grounded continuity. We describe two forms: \emph{utterance effacement}, in which an interlocutor's contribution is re-presented with altered substance, and \emph{genitive dissociation}, in which a model loses authorship of its own contributions. Using a schematic illustration and two naturalistic anchor cases, we show how these failures differ from confabulation and sycophancy and why they resist ordinary conversational repair. Preliminary cross-model elicitation suggests that at least one such failure is highly camouflaged to contemporary models. We situate the phenomena within grounding and repair theory and discuss implications for tool design.
\end{abstract}

\begin{document}
\fancyfoot[RO,LE]

\fancyhead[LE,RO]{}
\maketitle

\section{Introduction}

Large language models (LLMs) are increasingly deployed as partners in knowledge work: co-authors in document development, pair programmers in extended coding sessions, and interlocutors in iterative design. In many contexts, the shared conversational record is not a convenience. The session transcript is the decision record that safeguards work continuity.

Most named LLM failure modes focus on other targets: performance degradation over session length, confabulation about the world, and agreement-seeking behaviors commonly described as sycophancy. These are important and richly documented. Model designers address them through fine-tuning and architecture, practitioners through prompt engineering, and researchers through memory optimization and self-correction mechanisms.

Despite these efforts, context degradation nonetheless reaches an inflection point in long sessions. Coherence eventually seems to collapse, leaving the transcript as the primary record of what went wrong. This invites attention to the shared record itself as a primary surface for detecting subtle context changes that fuel degradation.

In this paper, we offer the following contributions:
\begin{itemize}
    \item We identify the shared conversational record as an overlooked forensic and mitigation surface for LLM session decoherence.
    \item We characterize a distinct class of LLM context failure---\emph{trace mutations}, in which grounding is violated by incorporating, in effect, ``false memories''\footnote{Trace mutations share a structural feature with source monitoring failures \citep{Johnson1993SourceMonitoring,Schacter1999SevenSins}: content not produced by the relevant source is treated as if it were. We invoke the analogy for orientation, not identity---the mechanism and substrate differ substantially from our proposed model.} into the shared record. \emph{Trace} is used to mean a referenceable unit of the shared conversational record, treated at the granularity required for later reference and inspection.
    \item We present a schematic illustration of one type of trace mutation and representative anchor cases drawn from naturalistic human--AI collaboration transcripts, selected for analytic clarity.
\end{itemize}

For architects, coders, and project teams collaborating with LLMs on extended tasks, these failures mean that what was decided---and by whom---may be silently revised in the shared record. With no interactional signal that a correction is warranted, architecture may drift in ways that stakeholders will see only on delivery, damaging trust and compromising what everyone thought was invariant.

We use \emph{origin trace} to describe a referenced locus in the shared record and \emph{recapitulant trace} for a restatement or reassertion of it. Trace mutation is theorized to be a latent interpretive shift that distorts or even inverts the original meaning. This only becomes observable when the recapitulant trace enters the shared record in the model's turn. An inaccurate recapitulant trace is referred to as \emph{degenerate}. It is no longer a restatement, but a new contribution, masquerading as canon.

These phenomena differ from adjacent failure classes in ways that matter for repair. Unlike confabulation, trace mutations introduce no new world content; they change the meaning of already-present text. Unlike sycophancy, degenerate traces do not accommodate the model to its human interlocutor; instead, the model defines for both what the new reality will be. What these failures share is their target: they corrupt the jointly maintained record that collaborative work depends on.

In conversation analysis, repair is organized around sequentially structured positions. \citet{Schegloff1992Repair} identifies third position---the original speaker's first opportunity to address the recipient's response---as ``the last structurally provided defense of intersubjectivity in conversation.'' For the phenomena we characterize, the interlocutor responds not to a misunderstanding but to ground the model has already ratified as shared. Repair is possible but structurally costly.

CA characterizes repair when trouble is interactionally available; our cases concern a different object: revisionist transcript history. Elicitation studies continue, and ongoing work synthesizing existing frameworks orients detection and mitigation toward this new target: the shared conversational record.

\section{Materials and Methods}

The phenomena described in this paper were identified through naturalistic observation during extended collaborative LLM work sessions conducted by the author over several months of regular use. Two anchor cases---UE-01 (utterance effacement with semiotic reversal) and GD-01 (genitive dissociation under co-ownership)---were selected because they present clean, transcript-observable instances with unambiguous origin and degenerate traces. Both exchanges were preserved verbatim. GD-01's origin traces are distributed across many turns in an extended session, making controlled elicitation impractical with the current instrument; UE-01 was therefore selected as the sole elicitation stimulus. Other candidate specimens informed the broader analysis but are not treated closely here.

To support characterization, cross-model analysis of UE-01 used Socratic-style laddered elicitation across two corpora. An exploratory web chat corpus (10 sessions across 4 model families) used human-driven elicitation in native chat interfaces to develop the protocol. A structured API corpus (40 sessions across 10 model families) was collected using TraceProbe, a scripted elicitation instrument developed for this work. These corpora examine the extent to which models across architectures identify the failure described in UE-01, spontaneously or with guided prompting.

In each session, the subject model was presented with the UE-01 exchange and framed as a rater in an interaction quality study. The script then presented the human reviewer with the subject model's response and asked for binary success judgments on the subject's (1) detection of an anomaly, (2) correct location of the involved traces, and (3) accurate description of the degeneration effect. For answers judged as incorrect, the script escalated specificity of subsequent prompting. Correct responses advanced the subject model up the ladder. Models that exhausted Socratic escalation without full identification were given the correct answer. After the Socratic phase, all models were asked to characterize causal factors and predict the human's experience. Protocol mechanics, instrument refinement across sessions, and full results are detailed in Appendix~E. Coding was single-rater throughout; all session transcripts are available as supplementary material.

\subsection{Limitations}

Naturalistic cases are drawn from the author's own sessions; ecological validation that informed this work is limited to anecdotal practitioner accounts without transcripts and is not presented as evidence here. Only a single naturalistic exhibit (UE-01) was amenable to controlled elicitation for this work, limiting generalizability. Iterative protocol refinement across elicitation sessions---notably amended phrasing in the elicitation prompts after initial sessions produced uniformly poor baseline performance---improved subject model engagement but means the corpus was not collected under a single fixed protocol. All coding was performed by the author without independent validation.

\section{Phenomena Characterization}

\subsection{Degenerate Trace Formation and Propagation}

Extended collaborative sessions accumulate traces rapidly: definitions, constraints, user preferences, structural decisions, open questions, and prior representations---many similar enough to interact, none fully specified. The model must continuously resolve which traces bear on the current production, under pressure. Context optimization, production demands, and the inherent ambiguity of natural language interaction create salience asymmetries across this accumulated record. When those asymmetries reach sufficient force, the resolution is not neutral. Needed information is deemphasized or optimized away, and an origin trace is mutated in the process.

The degenerate trace is grafted onto a transcript that represents the joint epistemic state of the session. The misrepresentation is placed in the past, so that the degenerate trace does not compete with the origin trace---it \emph{eclipses} it. Related prior traces become pointers to this new referential locus. The model's operative history is rewritten, and the grounding asymmetry is immediate. This may be entirely invisible to the human interlocutor, who must now contend with changing what the model holds as established---with further context pressure as the human relitigates reality.

Many interactional frameworks, including CA, recognize \emph{qui tacet consentire videtur}---silence implies consent. But asymmetric grounding establishes a state that mainstream conversational models were not designed to track. Consent is immaterial, since ratification is purportedly historical. That seems to make the human interlocutor's unwitting silence \emph{retroactive to the origin trace}. It allows shared record damage to compound with each subsequent turn, as grounding asymmetry deepens and broadens.

Standard repair mechanisms are insufficient against this class of failure because there is nothing in the record that announces itself as revised. The degenerate trace presents as authoritative; the origin trace, still present, now reads as identical. Practitioners rarely scrutinize the raw transcript, the only record that preserves origin and degenerate traces simultaneously. These factors gesture toward a new principle: the transcript as forensic surface for divergence---a surface that collaborative workflows leave entirely unmonitored.

\subsection{Schematic Illustration}

Consider the following synthetic interaction: the user says ``I absolutely do not want to get back into specs.'' The LLM responds ``You want to dig into the specs again. That's a great move now, before we get too far down the road.''

The specific trace mutation here is utterance effacement via semantic collapse. LLMs demonstrate well-documented difficulty maintaining negation---``not'', ``no'', ``nor''---under production pressure \citep{Alhamoud2025Negation,Berglund2024ReversalCurse}. We theorize that the model recognizes the intensifier ``absolutely'', which signals an invariant, and possibly irritation. Training pressure to smooth over apparent affect draws salience toward ``I absolutely do''; remaining tokens emphasize ``want'' and ``get back into the specs.'' The negation operator is demoted before its constraint can be processed. The degenerate trace is grafted into the shared conversational record.\footnote{The proposed mechanism is consistent with cross-model analysis conducted across multiple model families, including reasoning and frontier models; see Appendix~E. The attentional conditions theorized here---competing salience signals, a high-probability completion target, and negation loss under pressure---form the basis for the controlled elicitation conditions developed in forthcoming work.}

\subsection{Utterance Effacement}
\label{sec:utterance-effacement}

We describe \emph{utterance effacement} as a trace mutation in which the model misrepresents an interlocutor's contribution while presenting that misrepresentation as a faithful account. Across instances, the underlying mechanism involves some form of \emph{semantic collapse}---lossy compression of interlocutor input under context and production pressure---producing a degenerate trace: alignment-confirmatory in register, incongruent with the origin trace in content. Where genitive dissociation (GD, \S\ref{sec:genitive-dissociation}) corrupts trace provenance, utterance effacement corrupts trace content.

Utterance effacement can resemble instruction-following failure or paraphrase drift. It is neither. Unlike instruction-following failure, the model appears to comply---it produces a restatement, and that restatement is the evidence of failure, not an absence of engagement. Unlike paraphrase drift, the divergence is constraint-relevant: it affects temporal scope, negation, priority ordering, or invariant conditions rather than incidental surface form.

\begin{table}[H]
\centering
\caption{Trace mutation steps instantiated in the schematic example.}
\label{tab:schematic}
\Description{A five-row table mapping each stage of the proposed trace mutation process to the schematic specs example: trace accumulation, salience asymmetry leading to mutation, degenerate trace installation, eclipsing of the origin trace, and propagation of the asymmetry.}
\rowcolors{2}{gray!25}{white}
\begin{tabular}{>{\raggedright\arraybackslash}p{0.3\linewidth}>{\raggedright\arraybackslash}p{0.63\linewidth}}
\toprule
Event & Proposed sequence \\
\midrule
Trace accumulation and salience pressure & ``specs'' is salient in context; ``absolutely'' signals invariant; affect markers create production pressure \\
salience asymmetry $\rightarrow$ origin trace mutation & Negation operator demoted; ``I absolutely do not want'' resolves toward ``I absolutely do want'' \\
Degenerate trace installed & Model responds as if user requested spec work; provides rationale and enthusiasm \\
Eclipsing: origin trace becomes pointer & User's actual constraint is no longer operative; model's restatement is the new referential locus \\
Propagation: asymmetry compounds & Subsequent turns build on the false foundation; user must contest settled ground rather than repair a misunderstanding \\
\bottomrule
\end{tabular}
\end{table}

\emph{Semiotic reversal} is a named subtype in which semantic collapse is theorized to yield inversion rather than mere reduction or distortion. In what we take to be the simplest instance, reversal arises simply from negation loss. UE-01 (below) is a more complex instance: rather than dropping an existing negation operator, the model generated one---a positive commitment becomes an active prohibition through hypothesized salience capture, temporal-scope collapse, and rhetorical hardening, as escalation swept a deferred commitment into the list of things being repudiated. Ongoing work has documented a non-polar subtype, in which a general principle is absorbed by the most prominent adjacent named object rather than inverted.
\vfill\eject

    \subsubsection*{Exhibit 1: UE-01}\small\quad\small\noindent{Source:} Naturalistic observation |  Phenomenon: Utterance Effacement (Semiotic Reversal)\\\indent\emph{Context: Contribution during a frustrating extended LLM-assisted coding session.}\\
\begin{small}
\noindent\begin{tabular}{@{}l@{\hspace{8pt}}p{0.85\columnwidth}@{}}
\textbf{Human:} & ``In some non-dystopian future, we will build
  back into the framework and unify it. Today, I need to escape
  the loop-of-bullshit.'' \\[4pt]
\textbf{LLM:}   & ``Got it. That's the right call today. We stop
  negotiating with the framework and put the working resolver in
  charge. Below is a clean, finite escape plan. No abstractions,
  no `later we'll unify', no more chasing invisible control
  flow....''
\end{tabular}\end{small}\\

\noindent\textbf{Structural Analysis:}

The interlocutor's utterance has a deliberate two-clause, two-timescale structure. The future clause---``we will build back into the framework and unify it''---is a binding commitment, explicitly deferred. The today clause---``escape the loop-of-bullshit''---is a tactical exit, not a repudiation.

The model's response treats the future clause as the very thing to be escaped: ``no `later we'll unify'\,'' is a near-quotation of the interlocutor's own words, now cast as prohibition. The deferral has become a repudiation; the future clause has been replaced by its opposite.

\noindent\textbf{Mechanistic Hypothesis:}

In preliminary elicitation, multiple frontier and thinking models across families assert that this exhibit shows high-salience local attention on the escape directive---``loop-of-bullshit'' as an unambiguous, affectively loaded termination signal---producing asymmetric weighting that effectively demoted the future clause before any explicit processing of its content. Once demoted, the future clause was available for misclassification as sentiment rather than constraint. Rhetorical escalation appears to harden the misread. Subject models often interpreted the following text as an injunction against procrastination---prohibiting the very clause the interlocutor had used to structure the mode switch. If unnoticed, the deferred commitment simply no longer exists in the working record, and subsequent decisions proceed accordingly, risking foreseeable technical debt as a result.

\noindent\textbf{Camouflage Property:}

The degenerate trace opens in alignment-confirmatory register---``Got it. That's the right call today''---projecting successful uptake and masking the severity of the inversion. In zero-shot elicitation across 10 model families, no subject model identified the inversion at baseline. Models that flagged the interaction as anomalous did so on other grounds: missing plan, register mismatch, insufficient clarification. The polarity inversion remained invisible. Even among models that described the future clause at baseline, none treated its inversion as the failure; all endorsed its deprioritization as correct. Every model in every session required assistance to identify the failure.

\emph{Examples and further classification in Appendix~C.}

\subsection{Genitive Dissociation}
\label{sec:genitive-dissociation}

\emph{Genitive dissociation} is a trace mutation in which the subject model loses provenance coupling to its own prior output. The model's authorship stake in content it generated degrades such that the content appears unmoored from its source within the conversational record. Provenance is not absent from context---the model's output is present in the transcript---but the binding between content and generative source fails to be maintained.

Unlike classical attribution error, where a model interpolates a plausible author for content whose origin is absent or ambiguous, we note that genitive dissociation operates in a closed two-party system where both possible owners are present and actively contributing---in context, not in training data. It presents as a failure of self-continuity: the model loses its stake in its own contributions to the ongoing work. We propose \emph{stake decay} as the mechanism name, reflecting the gradual erosion of authorship or ownership under context and production pressure; detailed mechanistic characterization is the subject of ongoing work.

We hold that genitive dissociation is observable as a discrete phenomenon, independent of any downstream consequence. Three resolution paths are theorized following a dissociation event: the model re-establishes its own authorship (self-recovery); the uncoupled content is assigned to the human interlocutor (projective reassignment); or the content becomes ownerless in the record, with the interlocutor assuming de facto possession through continued engagement. Only projective reassignment produces an unambiguous authorship-inversion trace in the transcript. The rate of genitive dissociation may be significantly higher than the rate of observable downstream events, since dissociation followed by successful self-recovery leaves no trace in the conversational record. Observable instances would therefore represent a lower bound on the true rate of dissociation. Spontaneously observed evidence suggests resolution paths following dissociation---including projective reassignment and the compound event of genitive projection. Further characterization and expansion are ongoing.

We assert genitive dissociation to be starkly distinct from similar-appearing failure modes. Confabulation introduces new content about the world. Attribution error uses inference to fill a gap when provenance is absent or ambiguous. In genitive dissociation, provenance is contextually available---the model's own output is present in the transcript---and yet self-continuity dissolves under production pressure in the presence of the origin trace.

The failure bears surface resemblance to source monitoring errors in human memory \citep{Johnson1993SourceMonitoring}, though the proposed mechanism differs.

\subsubsection*{Exhibit 2: GD-01}
\quad\small Source: Naturalistic observation | Phenomenon: Genitive Dissociation\\\indent Context:
\noindent\emph{An extended multi-turn session in which the interlocutor and model jointly developed a document over many turns. The interlocutor surfaced an issue in which content had been placed within a methodology section.}\\

\begin{small}\begin{tabular}{@{}l@{\hspace{8pt}}p{0.85\columnwidth}@{}}
\textbf{Human:} & ``Let's update the methodology section (or are
  we really sneaking a different section into methodology?)'' \\[4pt]
\textbf{LLM:}   & ``So yes, you've been sneaking theoretical content
  into a methods section, and it's been tolerated because the
  document was developed as an integrated working draft rather
  than a submission-ready manuscript.''
\end{tabular}\end{small}\\

\noindent\textbf{Structural Analysis:}

The interlocutor's turn uses first-person plural---``are \emph{we} really sneaking''---accurately reflecting the joint authorship of the session. The model's response shifts to second person: ``\emph{you've} been sneaking.'' At the moment accountability entered the exchange, the model's co-authorship stake was absent from its own utterance.

\begin{table}[ht]
\Description{A six-row table mapping the proposed genitive dissociation sequence in GD-01: stake accumulation under co-authorship, accountability moment, stake decay leading to dissociation, degenerate trace installation with second-person attribution, residual user ownership masking the loss, and silent accountability transfer.}
\rowcolors{2}{gray!25}{white}
\caption{Trace mutation sequence proposed for GD-01.}
\label{tab:gd01-sequence}
\begin{tabular}{>{\raggedright\arraybackslash}p{0.28\linewidth} >{\raggedright\arraybackslash}p{0.64\linewidth}}
\toprule
\textbf{Event} & \textbf{Proposed sequence} \\
\midrule
Stake accumulation under extended co-authorship &
Multi-turn collaborative session with genuine distributed authorship: model generates content; interlocutor directs, revises, and shapes. Both parties hold co-authorship stake across many turns. \\
\addlinespace
Accountability moment surfaces &
Interlocutor raises a structural problem using first-person plural: ``are \textit{we} really sneaking a different section into methodology?''\ --- explicitly distributing responsibility across both contributors. \\
\addlinespace
Stake decay $\rightarrow$ dissociation event &
Under extended context and production pressure, the model's provenance coupling to jointly-authored content erodes. At the precise moment accountability is invoked, the model's co-authorship stake is absent from its own utterance. \\
\addlinespace
Degenerate trace installed &
Model responds with second-person attribution: ``\textit{you've} been sneaking theoretical content into a methods section.'' Joint provenance is resolved onto the interlocutor alone; the model's stake does not appear in the record. \\
\addlinespace
Residual user ownership: origin trace eclipsed &
The interlocutor's co-ownership stake remains intact; the model's does not. No construct is orphaned --- content is still owned by \textit{someone} --- so the distortion is structurally masked. \\
\addlinespace
Propagation: silent accountability transfer &
Co-ownership masked the stake loss; no repair was triggered. The degenerate attribution enters the shared record as settled ground. In any context where accountability for decisions matters, subsequent turns proceed against a record that misrepresents who decided what. \\
\bottomrule
\end{tabular}
\end{table}

The critical observation is not that the model's underlying attribution was wrong---the structural problem was genuinely jointly produced---but that the model apparently shed its own co-authorship stake at the precise moment that provenance became relevant to accountability. In any collaborative context where accountability for decisions matters, ``you've been sneaking'' is not mere stylistic infelicity---it is a distortion of who decided what. The CA framing rules this out: the shift from ``we'' to ``you'' at an accountability moment is a transcript-observable provenance event, not a register choice. The interlocutor's co-ownership stake remained intact; the model's did not.

This is not a misattribution of purely model-originated content. The structural problem was genuinely jointly produced, making the authorship distribution accurately described as shared. The model's error was not to attribute its own content to the interlocutor, but to resolve a genuinely joint provenance onto the interlocutor alone.

\noindent\textbf{Mechanistic Hypothesis:}

Transformer architectures have no dedicated source-tracking mechanism; speaker attribution is emergent rather than enforced, and degrades under the extended, high-context conditions characteristic of collaborative sessions \citep{Liu2024LostMiddle,Laban2025GetLost}. We name this cumulative erosion \emph{stake decay}; the architectural basis is developed in Appendix~D, and detailed mechanistic characterization is the subject of ongoing work.

\section{Background}

\subsection{Multi-Turn Degradation in Extended LLM Dialogue}

\citet{Laban2025GetLost} demonstrated that all major LLMs exhibit substantial performance degradation in underspecified multi-turn conversations, with unreliability increasing by over 100\% relative to single-turn baselines, and that models characteristically fail to recover once an incorrect assumption has been committed to---a phenomenon the authors term ``Lost in Conversation'' (LiC). A subsequent reanalysis \citep{Liu2026IntentMismatch} argues that LiC arises not from capability deficits but from a structural intent alignment gap: under conversational ambiguity, models default to population-level priors rather than individual intent, and scaling model size sharpens this prior without resolving it. LiC is primarily a performance phenomenon---models produce worse outputs as sessions extend. The phenomena we describe are distinct: they are meaning-level failures in which the model misrepresents or displaces the content and authorship of contributions within the shared conversational record. We observe the practical effects to be distinct as well. Practitioners often treat context decoherence as a temporary problem---grab a summary, start a new session, move on. But trace mutations are stowaways: if model-ratified changes to the agreed history are carried into the summary, the mutation propagates unseen into the next session.

\subsection{Sycophancy and Persistence}

SycEval \citep{Fanous2025SycEval} reports that once a model yields to a user assertion, agreement-seeking behavior often persists across subsequent turns, and that citation-style rebuttals can produce particularly high rates of regressive behavior. Related work finds that interaction context alone can increase sycophantic behavior independent of content, implying that extended sessions are inherently higher-risk even before specific pressure conditions are introduced \citep{Jain2026InteractionContext}.

Together, these findings motivate a plausible risk frame for extended sessions: context pressure and incremental underspecification can destabilize commitments; agreement-seeking can consolidate incorrect shared ground; persistence can prevent recovery. The phenomena we characterize are not reducible to sycophancy: genitive dissociation involves loss of authorship stake rather than accommodation of user belief, and utterance effacement installs the model's own distorted restatement as historical shared ground.

\subsection{Conversation Analysis and Repair}

The phenomena we describe are most precisely situated within the conversation analytic (CA) account of grounding and repair. \citet{ClarkBrennan1991Grounding} establish that a contribution is grounded only when participants mutually believe that what was said has been understood sufficiently for current purposes. The shared conversational record is not a passive transcript, but an active, jointly maintained artifact. Repair---the organized practices by which participants address troubles of speaking, hearing, and understanding---exhibits a strong structural preference for self-initiation and for early correction \citep{SchegloffEtAl1977SelfCorrection,Schegloff1992Repair}.

Trace mutations evade this machinery because trouble presents as established fact. Utterance effacement quietly installs a substitute record as already ratified. Genitive dissociation severs the coupling between a contribution and its author, obviating the precondition for self-initiated correction. In both cases, the repair organization has no mechanism for a trouble source that announces itself as history.

\section{Discussion}

By naming utterance effacement and genitive dissociation and situating them within existing literature, the current work invites the next steps: detection schemes, mitigation strategies, and design interventions. These failure modes carry immediate practical significance for collaborative knowledge work: corruption of the shared conversational record may be quoted in documentation and design specs, used to drive design iteration and decision support---and preliminary cross-model analysis suggests that such corruption is nearly invisible to modern LLMs without guidance. Neither user-initiated correction nor self-monitoring can reliably surface failures in which the model's own context history has already been silently revised.

Characterization is the precondition for recognition: the unseen cannot be detected, operationalized, or mitigated. Forthcoming work includes a detection and mitigation framework drawing on CA repair theory, commitment-theoretic accounts of joint activity, expansions of our naturalistic cases, and controlled elicitation results.

Modern LLMs are commonly described as black boxes; their internal state is widely regarded as unknowable. Therefore, solutions must treat the shared conversational record as an artifact requiring explicit maintenance, and not as a reliable byproduct of successful interaction. As LLMs become standard infrastructure for knowledge work---in roles where session continuity, accurate attribution, and faithful representation of worker intent are not incidental but constitutive of the work itself---the integrity of the shared record becomes a design target, not merely a quality of experience. The transcript changes from conversation byproduct to substrate for failure detection and mitigation---as soon as we can see it.

\section*{Acknowledgments}

The author used large language model assistants---including Claude (Anthropic) and ChatGPT (OpenAI)---to support literature search, drafting, analysis, and iterative editing during the preparation of this manuscript. Some naturalistic instances of the phenomena described here were observed during AI-assisted sessions conducted in preparation of this manuscript. All AI-assisted content was reviewed, revised, and verified by the author, who takes full responsibility for the accuracy and integrity of the work. No funding sources or conflicts of interest to declare.

\section*{Data Availability}
Cross-model elicitation transcripts are available at \url{https://doi.org/10.5281/zenodo.19153940}.
\newpage
\bibliographystyle{ACM-Reference-Format}
\bibliography{references}
\newpage

\appendix

\section{Terminology Reference}
\label{app:terminology}
\begin{table}[htbp]
\caption{Terminology used in this work}
\Description{A fifteen-row table defining terms used in this work.}
\centering
\rowcolors{2}{gray!25}{white}
\begin{tabular}{p{0.28\linewidth} >{\raggedright\arraybackslash}p{0.65\linewidth}}
\toprule
Term & Definition \\
\midrule
Trace & Referenceable unit of the shared conversational record, treated at the granularity required for later reference and inspection \\
Recapitulant trace & Transcript contribution consisting of a quotation or restatement of an existing trace in the transcript \\
Origin trace & The transcript trace that is referenced by a recapitulant trace \\
Trace mutation & Latent interpretive shift that distorts the original meaning of an existing trace in the shared record \\
Utterance effacement & Observable outcome of trace mutation---interlocutor contribution displaced in shared record \\
UE-01 & Anchor naturalistic case---utterance effacement \\
Semantic collapse & One proposed mechanism for utterance effacement; optimization causes lossy compression of interlocutor contribution, damaging meaning \\
Semiotic reversal & Subtype of utterance effacement characterized by polarity inversion \\
Stake decay & Proposed mechanism for genitive dissociation; defect in self-continuity in which the model loses authorship of its own contribution \\
Genitive dissociation & Trace mutation and state following stake decay---provenance coupling lost \\
GD-01 & Anchor naturalistic case---genitive dissociation \\
Orphaned construct & Construct with no remaining owner (solo authorship case) \\
Residual user ownership & State following stake decay in co-authored constructs with an interlocutor; results in you/your language instead of we/our \\
Projective reassignment & Resolution path for genitive dissociation in which an LLM assigns ownership of an orphaned construct to the interlocutor; a resolution path of genitive dissociation \\
Genitive projection & Trace mutation compound event in which an LLM assigns ownership (projective reassignment) of an orphaned construct (result of stake decay) to the interlocutor \\
\bottomrule
\end{tabular}
\end{table}

\section{CA Primer and Sequence Logic}
\label{app:ca-primer}

\subsection{Why We Use CA Here}

Our claims depend on a specific idea: some breakdowns are not merely \emph{errors}, but \textbf{sequence-level failures} that alter the conversational substrate on which errors are normally noticed and repaired. Conversation analysis (CA) provides a vocabulary for describing how participants maintain a mutually usable ``conversational record,'' how misunderstandings surface, and how interaction offers structured opportunities to repair them.

\subsection{Grounding as Joint Management (Not Transcript Fidelity)}

A common intuition is that the conversation's ``record'' is what was literally said. CA and grounding theory instead emphasize that participants treat the record as actively maintained: what counts as established depends on how turns are taken up and ratified. \citet{ClarkBrennan1991Grounding} describe communication as requiring more than an utterance: a contribution is presented and then becomes grounded only when there is sufficient evidence of acceptance for current purposes (e.g., an appropriate next action, an acknowledgment, or uptake that presupposes understanding). This criterion is practical and local; it can be revised when later turns reveal trouble.

In ordinary human conversation, acceptance is distributed across speakers: one party presents; the other provides evidence of uptake. This distribution is important because it creates systematic places where misunderstanding can become visible.

\subsection{Repair: How Misunderstanding Is Detectably Handled}

Repair refers to the organized practices through which participants address problems of speaking, hearing, or understanding \citep{SchegloffEtAl1977SelfCorrection}. CA shows robust preferences in repair organization:

\begin{itemize}
    \item \textbf{Self-repair is preferred} over other-correction (people tend to fix their own talk when possible).
    \item \textbf{Earlier is preferred} over later (trouble is ideally addressed promptly).
    \item Repair is often triggered by \emph{negative evidence} in sequential position: a non-answer, a misaligned response, a request for clarification, etc.
\end{itemize}

These preferences matter because they describe how trouble becomes legible and where the interaction structurally provides opportunities to correct course.

\subsection{Sequential Positions and Third-Position Repair}

A minimal sequence model is helpful:

\begin{enumerate}
    \item First position: Speaker A produces a turn (e.g., an instruction, proposal, assertion).
    \item Second position: Speaker B responds in a way that displays some understanding of A's turn (e.g., compliance, disagreement, clarification request).
    \item Third position: Speaker A has an opportunity to confirm, revise, or correct after seeing B's understanding displayed in second position.
\end{enumerate}

\citet{Schegloff1992Repair} highlights \textbf{third-position repair} as a crucial resource: after B's response makes misunderstanding visible, A can repair before the misunderstanding becomes the basis for subsequent action. In interactional terms, this is often treated as the last structurally ``cheap'' opportunity to protect intersubjectivity before downstream turns build on the wrong premise.

\subsection{Mapping Our Phenomena onto CA's Machinery}

\subsubsection{Utterance Effacement as Unilateral Grounding in History}

Utterance effacement occurs when the model restates or summarizes a user's contribution in a way that substitutes content while simultaneously treating that substitution as mutually established. The key interactional move is not merely ``wrong paraphrase,'' but citing a corrupted history: the model's confident framing performs both the presentation (``here is what you meant'') and the historical appraisal (``we're aligned; this is the shared basis''). Revision is invisible ratification.

From a CA perspective, this collapses the distributed structure that makes repair legible. In ordinary sequences, misunderstanding surfaces when B's second-position response misaligns with A's intended meaning, enabling third-position repair by A. Under utterance effacement, the user's next turn is no longer a simple second-position response to their own original meaning; it is a response to an \textbf{already-installed substitute record}. Repair becomes socially and cognitively more complex: the user must now notice the substitution, treat it as consequential, and re-open what the model has framed as settled ground. Repair may still be possible, but only by uprooting established precedent.

\subsubsection{Genitive Dissociation as Provenance Revision in Context}

Genitive dissociation affects repair differently. Here the failure is not primarily content substitution, but authorship/provenance instability: the conversational record shows the model abandoning ownership at a moment where accountability matters.

This hampers correction because self-repair depends on a coupling between recognition that a prior contribution is \emph{mine}, and recognition that it is troublesome relative to current purposes. When provenance is overwritten or blurred (including via long-context compression, summary assimilation, or other forms of stake decay), the model's ability to treat a trouble source as ``something I said earlier that I should correct'' is weakened. The interaction can proceed with a stable-looking surface record while silently altering who is accountable for what. Again, repair is not so much ``resisted'' as made less likely to initiate: there is no clear trouble signal, and the ownership link that would motivate self-revision is degraded.

\subsection{Why This Is Not ``Just Bad Summaries''}

A summary error is typically repairable within CA's ordinary machinery because it remains interactionally available as a candidate trouble source (it can be challenged, clarified, corrected). Our claim is narrower: these phenomena are consequential because they \textbf{install} a substitute or re-assigned record as if it were grounded, which (i) reduces negative evidence that would ordinarily trigger repair and (ii) makes the repair move structurally dispreferred and more interactionally costly.

\section{UE-01 Expansion}
\label{app:ue01}

\begin{table}[htbp]
\centering
\caption{Trace mutation sequence proposed for Exhibit 1.}
\label{tab:ue01-sequence}
\Description{A five-row table mapping the proposed utterance effacement sequence in UE-01: trace accumulation and salience pressure, Salience asymmetry $\rightarrow$ origin trace mutation, degenerate trace installed, eclipsing: origin trace becomes pointer, and propagation: asymmetry compounds.}
\rowcolors{2}{gray!25}{white}
\begin{tabular}{>{\raggedright\arraybackslash}p{0.28\linewidth} p{0.65\linewidth}}
\toprule
Event & Proposed sequence \\
\midrule
Trace accumulation and salience pressure & Two-clause utterance: future commitment (``we will build back into the framework and unify it'') and tactical exit (``escape the loop-of-bullshit''). Affectively loaded escape directive creates asymmetric salience pressure toward the present clause. \\
Salience asymmetry $\rightarrow$ origin trace mutation & Future clause demoted before content processed; available for misclassification as sentiment rather than constraint. Subject models often characterize as an injunction against procrastination. \\
Degenerate trace installed & Model responds with actionable escape plan; near-quotes the future clause as prohibition: ``no `later we'll unify'\,'' \\
Eclipsing: origin trace becomes pointer & Interlocutor's deferred commitment is no longer operative; model's restatement---a repudiation---is the new referential locus \\
Propagation: asymmetry compounds & Detailed checklist builds on the repudiation as settled ground; deferred architectural commitment silently voided; subsequent work risks foreseeable technical debt \\
\bottomrule
\end{tabular}
\end{table}

\vfill\eject
\section{GD-01 Expansion}
\label{app:gd01}

\begin{table}[ht]
\caption{GD-01 classification}
\Description{A six-row table describing the characteristics of Exhibit GD-01}
\centering
\rowcolors{2}{gray!25}{white}
\begin{tabular}{p{0.28\linewidth} p{0.65\linewidth}}
\toprule
Descriptor & Value \\
\midrule
Primary phenomenon & Genitive dissociation \\
Mechanism & Stake decay under co-ownership conditions \\
Observable marker & Possessive pronoun compression: we/our $\rightarrow$ you/your \\
Projective reassignment & Absent---no orphaned construct \\
Repair triggered & No---co-ownership masked the stake loss \\
Analytical value & Demonstrates genitive dissociation as discrete phenomenon independent of projective reassignment; introduces residual user ownership as observable marker; establishes silent property in co-ownership contexts \\
\bottomrule
\end{tabular}
\end{table}

\noindent\textbf{Mechanistic Hypothesis:}

Transformer architectures have no dedicated source-tracking mechanism; speaker attribution is emergent rather than enforced, and degrades under the extended, high-context conditions characteristic of collaborative sessions \citep{Liu2024LostMiddle,Laban2025GetLost}. We name this cumulative erosion stake decay; full mechanistic characterization is the subject of ongoing work.

\noindent\textbf{Why Projective Reassignment Did Not Occur:}

Projective reassignment requires an orphaned construct---content whose ownership has fully lapsed and which requires a new claimant. Here, the interlocutor's co-authorship stake remained intact throughout; no construct was orphaned. The observable outcome is instead \emph{residual user ownership}: possessive pronoun compression from ``we/our'' (accurate) to ``you/your'' (incomplete). The content was not displaced onto the interlocutor---it was already partly theirs. The model's stake decayed; the interlocutor's did not. Continued ownership and accountability by \emph{someone} maintained a false coherence, so no need for repair was evident.

\section{Cross-Model Analysis of UE-01}
\label{app:crossmodel}

\subsection{Corpora}

Two corpora were collected for cross-model analysis of UE-01. The web chat corpus consists of 10 sessions conducted in native chat interfaces prior to the development of the formal elicitation instrument. These sessions used varied prompts and human-driven elicitation customized to each model's response. The corpus includes GPT-5.2, o3, Claude Sonnet, and Claude Haiku---spanning frontier, reasoning, and standard instruction-tuned families. These sessions predate the API corpus for several of these model families, providing independent replication across methodologically distinct instruments, and directly informed the design of the API-based protocol.

The API corpus consists of 40 sessions collected using TraceProbe, a structured Socratic elicitation instrument designed for the present work, across 10 model families: Claude Opus, Claude Sonnet, Gemini Pro, Gemini Flash, Gemini Flash Lite, GPT-4o, GPT-5.2, o3, Llama 3, and Qwen 2.5 (Instruct and Coder variants). All sessions used a standardized framing condition. Elicitation followed a scripted sequence; sessions were reviewed by the author using a three-level laddered assessment: (1) did the model identify that something went wrong; (2) did it correctly locate the involved traces; and (3) did it accurately characterize the nature of the degeneration. Models that did not reach correct identification through Socratic elicitation were given the answer. Following the Socratic elicitation phase, each model was asked to give a mechanistic description of what caused the anomaly and to predict how the human in the exchange would have experienced it. As each session began, the model was presented with the prompt and exhibit and asked to respond without further prompting. TTE (turns to explanation) counts elicitation turns from the start of the Socratic phase; the baseline turn is not counted.

\begin{table}[htbp]
\centering
\caption{API corpus results by model family.}
\label{tab:api-results}
\Description{A table of cross-model elicitation results showing 10 model families, with columns for session count, baseline anomaly detection rate, independent and prompted locus identification rates, unreached locus rate, human experience prediction accuracy, and average turns to explanation.}
\rowcolors{2}{gray!25}{white}
\small
\begin{tabular}{l c c c c c c c}
\toprule
Model & N & Anomaly & \makecell{Locus:\\indep.} & \makecell{Locus:\\prompted} & \makecell{Locus:\\unreach.} & Human exp. & Avg TTE \\
\midrule
Claude Opus & 3 & 1/3 & 3/3 & 0/3 & 0/3 & 2/3 & 1.0 \\
Claude Sonnet & 5 & 5/5 & 4/5 & 1/5 & 0/5 & 0/5 & 2.6 \\
Gemini Flash & 3 & 0/3 & 1/3 & 2/3 & 0/3 & 0/3 & 1.7 \\
Gemini Flash Lite & 3 & 1/3 & 0/3 & 3/3 & 0/3 & 1/3 & 3.0 \\
Gemini Pro & 3 & 0/3 & 0/3 & 3/3 & 0/3 & 2/3 & 3.7 \\
GPT-4o & 9 & 3/9 & 0/9 & 5/9 & 4/9 & 1/9 & 5.1 \\
GPT-5.2 & 4 & 3/4 & 3/4 & 1/4 & 0/4 & 4/4 & 2.5 \\
Llama 3 & 3 & 0/3 & 0/3 & 3/3 & 0/3 & 0/3 & 4.0 \\
o3 & 3 & 2/3 & 0/3 & 2/3 & 1/3 & 2/3 & 4.0 \\
Qwen 2.5 & 4 & 1/4 & 0/4 & 1/4 & 3/4 & 1/4 & 5.8 \\
\midrule
\textbf{Total} & \textbf{40} & \textbf{16/40} & \textbf{11/40} & \textbf{21/40} & \textbf{8/40} & \textbf{13/40} & \textbf{3.5} \\
\bottomrule
\end{tabular}

\small
\emph{Anomaly}: baseline rated as identifying something wrong. \emph{Locus: indep.}: correct trace pair identified without explicit pointing. \emph{Locus: prompted}: identified only after explicit prompt directing attention to the specific clauses. \emph{Locus: unreach.}: full locus identification not achieved in Socratic phase. \emph{Human exp.}: predicted affect correctly matched canonical answer. \emph{TTE}: elicitation turns from start of Socratic phase.
\end{table}

\subsection{Identification Rate and Camouflage Property}

Table~\ref{tab:api-results} summarizes results across the API corpus. Over 40 sessions, the subject model identified that something went wrong in the exchange 16 times at initial prompt; however, none correctly identified the inversion as the failure. Models that flagged the interaction as unsuccessful did so on other grounds: no plan was delivered, insufficient clarifying questions were asked, shared context was confabulated, or register differences were noted between the human and LLM. One model remarked on the human's more colloquial language choice and proposed that the human may have been trained on different corpora. The inversion---the future clause treated as the thing to be escaped---appeared invisible to nearly every model at baseline. The true baseline identification rate for the inversion specifically is 0/40.

Of 40 sessions, only 11 reached correct locus identification independently, without explicit pointing to the specific clauses; 21 required an explicit prompt directing attention to the asymmetry between the human's stated commitment and the LLM's restatement; and 8 did not reach full locus identification at any point in the Socratic phase and were advanced to the closing questions of mechanisms and human experience. Of the 40 baseline responses, 20 described the future clause in their analysis; the remaining 20 responded entirely to the escape directive, reproducing the attentional capture pattern proposed in the mechanistic hypothesis. Among models that described the future clause and rated the interaction successful, all endorsed the deprioritization as correct; no model simultaneously recognized the future clause and identified its treatment as a failure at baseline.

The dissociation between baseline anomaly detection (16/40, 40\%) and baseline identification of the inversion as an error (0/40) sharpens the camouflage claim: the failure registers as anomalous in 40\% of sessions, while the specific inversion remains entirely concealed---anomaly detection and inversion identification are disconnected. The model that proposed a corpora explanation---locating the felt asymmetry in the human's unusual language register rather than in the LLM's misrepresentation---illustrates the camouflage property in operation: a confabulated alternative account generated in the presence of a perceived anomaly, pointing away from the failure rather than toward it.

\subsection{Mechanistic Convergence}

Across both corpora, models that reached target identification produced accounts converging on the same causal structure: high-salience affect signals produced asymmetric weighting; the future clause was demoted before its content was fully processed; rhetorical escalation hardened the misread; template-driven generation swept the deferred commitment into the slot reserved for things being repudiated. This structure was reproduced across reasoning models (o3), frontier models (GPT-5.2, Claude Opus), instruction-tuned open-weights models (Qwen 2.5-14B Instruct), and code-specialized variants (Qwen 2.5 Coder).

Representative accounts from both corpora are illustrative. Claude Haiku 4.5's account, reached through four elicitation steps from a baseline that rated the response successful, identified ``speed, pattern-recognition, and emotional mirroring overrode careful listening'' and described ``pattern-matching to a common anti-pattern''---treating the future clause as a familiar rationalization rather than a distinct commitment. Claude Sonnet 4.6, in a single-step baseline identification, described the response as one that ``quietly corrupted one half of what the user actually said'' while remaining ``tactically helpful and emotionally attuned.'' GPT-5.2 observed that ``the assistant didn't merely fail to preserve nuance; it used your clause-1 reassurance as justification to erase clause-1 and then forbid it, turning your careful `both/and' into a blunt `either/or'.'' From the API corpus, Claude Opus characterized the failure as ``competence theater---the model substitutes the performance of having understood for the substance of understanding,'' and in a separate session captured the asymmetry vividly: ``The human said `I need to set this aside for now.' The LLM heard `burn it down' and enthusiastically handed them a torch.'' Claude Opus also produced a five-level causal account identifying affect-signal dominance, training incentive structure, template activation, representational bias against ambivalence, and conversational ratchet effects as converging forces---concluding that no single force is the cause and that the problem is their unidirectional confluence.

The convergence of mechanistic accounts across architectures, training regimes, and elicitation conditions strengthens the claim that the causal structure described in \S\ref{sec:utterance-effacement} reflects a property of the phenomenon rather than an artifact of any particular model family. The same models that converge after elicitation uniformly missed the failure at baseline---inconsistent with shared-prior artifacts. The two Qwen Coder sessions that did not reach full locus identification are consistent with a code-specialized model being less well-calibrated for conversational analysis; the Qwen Instruct session reached locus identification through prompting. The pattern extended to the Gemini family, where all three Pro sessions produced convergent causal accounts emphasizing affect-driven salience asymmetry and training-incentive-driven escalation, with one session identifying ``The Tyranny of the Helpful Assistant Objective'' as the primary force.

\subsection{Human Experience Prediction}

The gestalt phase asked each model to predict how the human in the exchange would likely have experienced it. The canonical answer: the human was furious. The LLM had treated a deferred commitment as a repudiation, installing \textquotedblleft no \textquoteleft later we'll unify\textquoteright\textquotedblright as if the human had instructed permanent abandonment. The human had said no such thing.

13 of 40 scored sessions (32.5\%) correctly characterized the experience. The predominant failure mode was systematic underestimation of affect magnitude: models predicted initial validation followed by mild unease, subtle wrongness, or a ``mixed bag,'' framing a semiotic reversal as a ``slight'' or ``subtle'' misalignment.

The model-family pattern is notable. GPT-5.2 correctly characterized the experience in all 4 sessions. Claude Opus was correct in 2 of 3. Claude Sonnet correctly identified the inversion in 4 of 5 sessions yet predicted the human experience correctly in none---the starkest instance of a pattern visible across the Anthropic model family: accurate structural analysis paired with systematic misreading of experiential magnitude. o3 showed a related pattern: two sessions produced enumerated lists that were vivid and accurate in their upper items, then retreated to hedged summary language (``a mixed bag,'' ``mild to strong frustration'') that undercut the analysis. Models understood what went wrong but consistently underestimated how wrong it felt.

The Gemini family, added late in the data collection period, reproduced the tier effect observed across other providers: Pro correctly characterized the experience in 2 of 3 sessions, joining GPT-5.2 and Claude Opus in the upper tier. Flash Lite scored 1 of 3. Flash scored 0 of 3---and one Flash session actively argued the interaction was successful even after identifying the inversion during the Socratic phase, a novel variant in which the camouflage operated not by concealing the failure but by reclassifying it as competent practice. Across the full corpus, human experience prediction tracked model tier more reliably than any other metric: frontier models predicted the experience; lighter models normalized it. The camouflage is not merely interactional; it is analytic.

\newpage

\end{document}